\documentstyle[12pt,preprint,eqsecnum,aps]{revtex}
\begin{document}
\title{$\phi$-Meson Mass Modification in Heavy-Ion Collisions} 
\author{R. S. Bhalerao$^1$, S. K. Gupta$^2$ and P. Shukla$^2$}
\address{$^1$Theoretical Physics Group,
Tata Institute of Fundamental Research\\
Homi Bhabha Road, Colaba, Mumbai 400 005, India}
\address{$^2$Nuclear Physics Division,
Bhabha Atomic Research Centre\\
Trombay, Mumbai 400 085, India}
\maketitle
\begin{abstract}
We recently presented a method of analyzing invariant-mass spectra of
kaon pairs resulting from decay of $\phi$ mesons produced in
high-energy heavy-ion collisions. It can be used to extract the shifts
in the mass and the width ($\Delta M$ and $\Delta \Gamma$) of the
$\phi$ mesons when they are inside the dense matter formed in these
collisions. We have now performed a Monte-Carlo simulation of this
process. We illustrate our method with the help of available
preliminary data. Extracted value of $\Delta M$ is significantly
larger than that obtained with an earlier method. Our results are
consistent with the experimentally observed $p_T$ dependence of the
mass shift.
\end{abstract}

\medskip
\noindent{\bf {1.~ Introduction}}
\medskip

Relativistic heavy-ion collisions and the consequent formation of a
dense and possibly thermalized blob of matter provides a means to test
the theoretical prediction of (partial) restoration of chiral symmetry,
which would manifest itself in a downward shift in the mass of a
hadron so long as it is inside the blob. Here we outline our 
work$^1$ regarding the $\phi$-meson mass and describe results
of a Monte-Carlo simulation that we have performed more recently.

Recently E-802 collaboration at AGS (BNL) has reported preliminary
results on the shift in the mass of the $\phi$ mesons produced in
central $Si + Au$ collisions at 14.6 A GeV/c.$^{2,3}$ It was found
that in the events with the highest multiplicity (top 2$\%$
target-multiplicity-array cut) the mass drops by $2.3 \pm 0.9 \pm
0.1$ MeV compared to the free-space value. For the $\phi$ mesons with
$p_T < 1.25$ GeV/c, the shift was even more ($3.3 \pm 1.0 \pm 0.1$
MeV), whereas for $p_T > 1.25$ GeV/c, there was no apparent shift. No
numbers were reported for the shift in the {\it decay width} of the
$\phi$ mesons. However, from the confidence contours for 1, 2 and 3
standard deviations in the observed mass versus width plot given in
Ref. 3, it appears that the central value of the width is higher than
the free-space value by about 0.78 MeV.

\medskip
\noindent{\bf {2.~ Method}}
\medskip

We present a different method of analyzing the
invariant-mass spectra of kaon pairs. We then use the data in
Refs. [2-3] only to illustrate our method. We find that our method
yields a shift in the $\phi$-meson mass which is significantly larger
than the values quoted above. We caution the reader that since these
data are preliminary, our numerical results are obviously subject to
change. {\it However, our main point is the method presented here
which is independent of whether the published data$^{2,3}$ are
eventually confirmed or not.} We think the present method has
relevance to the analysis of {\it future data} on $\phi$ production.
This acquires added importance in view of a similar experiment having
been performed at SPS (CERN). 

Let $M_0$ and $M_1$ be the rest masses and let $\Gamma_0$ and $\Gamma_1$
be the widths of the $\phi$ meson in the free space and in the dense
medium, respectively. We define the shifts as
$\Delta M = M_0 - M_1$ and $\Delta \Gamma = \Gamma_1 -\Gamma_0,$
so that both are positive when the mass drops and the width increases
with respect to their free-space values.

In Ref. 3, the invariant-mass spectrum of the $K^+K^-$ pairs was
fitted by a function consisting of a background term and a
relativistic Breit-Wigner (BW) resonant term convoluted with a
Gaussian experimental mass resolution function; see Fig. 1. This
procedure yielded the values of $\Delta M$ and $\Delta \Gamma$ given
above. However, are these values obtained by fitting a {\it single} BW
term to the (background-subtracted) data correct? We think they are
not.

Since the mean lifetime of $\phi$ in its rest frame is about 45 fm/c,
a majority of $\phi$'s are expected to decay long after the dense
medium in which they were produced has ceased to exist. That is, they
will decay essentially in free space ($\Delta M = 0 = \Delta
\Gamma$). The rest of the $\phi$'s will decay while still inside
the dense medium. Hence a better procedure would be to fit the
background-subtracted data with two instead of one BW terms, one
unshifted and the other shifted, added with appropriate weights.
This we now proceed to do.

We work in the centre-of-mass frame of the dense medium.
Let $f$ be the fraction of $\phi$'s decaying {\it inside} the medium;
then $(1-f)$ is the fraction decaying in free space. 
The unshifted BW has mass $M_0$, width $\Gamma_0$ and
weight $(1-f)$, and the shifted BW has
mass $M_1$, width $\Gamma_1$ and weight $f$. We reanalyze the 
background-subtracted data$^3$ by using:
\begin{equation}
{dN_{K^+K^-} \over dM} = (1-f)~ BW_c(M, M_0, \Gamma_0) 
+ f~ BW_c(M, M_1, \Gamma_1).
\eqnum{1}
\end{equation}
Here $BW_c$ denotes the relativistic BW term
convoluted with a Gaussian experimental mass resolution function;
see Ref. [1] for details. If the effect of the
time dilation on the lifetime of a $\phi$ is taken into account, 
the fraction $f$ can be shown$^1$ to be given by 
\begin{equation}
f(\Delta M, \Delta \Gamma) = 1 -\exp(-M_1 \Gamma_1 d/\sqrt{M_T^2~ {\rm
cosh}^2y_{cm} - M_1^2}), \eqnum{2}
\end{equation}
where $M_T$ and $y_{cm}$ denote, respectively, transverse mass and
centre-of-mass rapidity of the $\phi$ traversing the medium and $d$ is
the distance traversed. We have used a Monte-Carlo simulation
procedure to calculate $f$ by averaging over these three variables.
We generated $\phi$'s according to the experimental distributions in
$M_T$ and $y_{cm}$. The distance $d$ was also sampled by assuming a
cylindrical geometry for the medium. More details of this will be
published elsewhere.

\medskip
\noindent{\bf {3.~ Results}}
\medskip

Results of a least-squares fit to the same experimental data as in
Fig. 1, with two instead of one BW terms, are shown in Fig. 2. The
two fitted parameters are $\Delta M = 6.1 \pm 1.8 $ MeV and $\Delta
\Gamma = 4.2 \pm 3.9 $ MeV. The resultant value of $f$ is 0.35. Note
that the above value of the mass shift is significantly larger than
that obtained with a one-BW fit.

It is evident from Eq. (2) that as $p_T$ increases the fraction $f$
decreases. This means a larger fraction of $\phi$'s decay in free
space and hence the mass shift decreases. This is exactly what has
been observed experimentally.$^3$

\medskip
\noindent{\bf {References}}
\medskip

1.  R. S. Bhalerao and S. K. Gupta, Mod. Phys. Lett. A 12 (1997) 127.

2.  Y. Akiba et al., Nucl. Phys. A 590 (1995) 179c.

3.  Y. Wang, Nucl. Phys. A 590 (1995) 539c.

\vfill

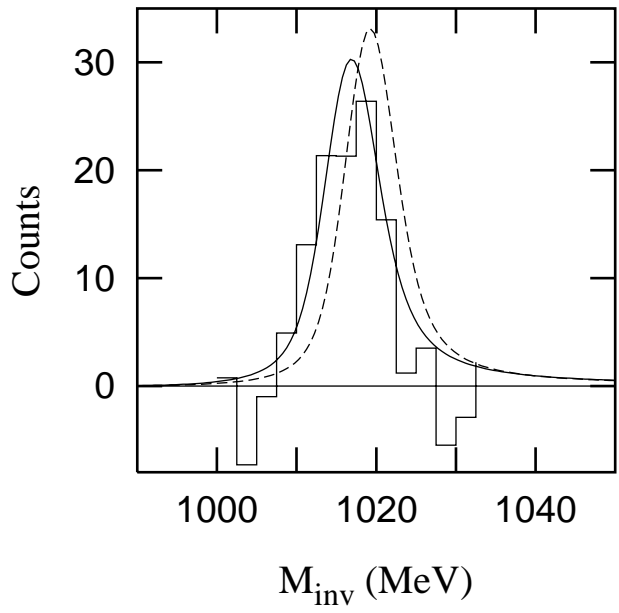
Fig. 1: The background-subtracted invariant-mass spectrum for the
$K^+K^-$ pairs. The histogram represents data from Ref. 3. The two
curves represent two convoluted BW terms. The dashed curve corresponds
to $\Delta M = 0 = \Delta \Gamma$, and the solid curve to $\Delta M =
2.3$ MeV and $\Delta \Gamma = 0.78$ MeV.

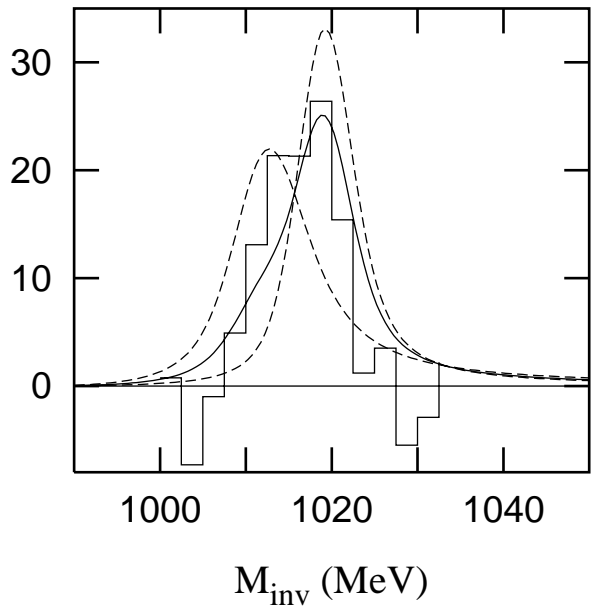
Fig. 2: The two dashed curves represent two convoluted BW terms, one
unshifted and the other shifted. The solid curve corresponds to their
weighted sum as in Eq. (1).

\end{document}